\documentclass[twocolumn, aps, prl, superscriptaddress,showpacs, floatfix]{revtex4}

\usepackage{multirow}
\usepackage{epsfig}
\usepackage{amsmath}

\usepackage{subfigure}
\usepackage{latexsym}
\usepackage{feynmp}

\usepackage[normalem]{ulem}  
\usepackage[dvips]{color} 

\renewcommand\sout{\bgroup \color{red} \ULdepth=-.5ex \ULset}

\begin{document}

\title{Freeze-out conditions for production of resonances, hadronic
molecules, and light nuclei}

\author{Sungtae Cho}
\affiliation{Division of Science Education, Kangwon National
University, Chuncheon 200-701, Korea}
\author{Taesoo Song}
\affiliation{Frankfurt Institute for Advanced Studies and
Institute for Theoretical Physics, Johann Wolfgang Goethe
Universitat, Frankfurt am Main, Germany}
\author{Su Houng Lee}
\affiliation{Department of Physics and Institute of Physics and Applied Physics, Yonsei
University, Seoul 120-749, Korea}

\begin{abstract}
We investigate the freeze-out conditions of a particle in an
expanding system of interacting particles in order to understand
the productions of resonances, hadronic molecules and light nuclei
in heavy ion collisions. Applying the kinetic freeze-out condition
with explicit hydrodynamic calculations for the expanding hadronic
phase to the daughter particles of $K^*$ mesons, we find that the
larger suppression of the yield ratio of $K^*/K$ at LHC than at
RHIC compared to the expectations from the statistical
hadronization model based on chemical freeze-out parameters
reflects the lower kinetic freeze-out temperature at LHC than at
RHIC. Furthermore, we point out that for the light nuclei or
hadronic molecules that are bound, the yields are affected by the
freeze-out condition of the respective particle in the hadronic
matter, which leads to the observation that the deuteron
production yields are independent of the size of deuteron, and
depend only on the number of ground state constituents.
\end{abstract}

\pacs{25.75.Dw, 25.75.-q}


\maketitle

Relativistic heavy ion collision experiments have enriched our
understanding on the properties of quantum chromodynamic (QCD)
matter at high temperature and densities \cite{Adams:2005dq,
Adcox:2004mh, Gyulassy:2004zy}. One of the important experimental
objectives in relativistic heavy ion collision experiments is
confirming the occurrence of the phase transition between a system
composed of free quarks and gluons, the so-called quark-gluon
plasma (QGP), and the hadronic matter that was predicted by
Lattice QCD calculations \cite{Gupta:2011wh}. In that respect, the
large numbers of various kinds of hadronic particles produced at
the phase boundary have been useful in elucidating the entire
process of heavy ion collisions, since production yields of hadron
are affected by conditions at each step of the system evolution.

The systematic studies on the hadron production yields using the
statistical hadronization model have led to the identification of
conditions at the phase boundary; the phase transition temperature
and the baryon chemical potential \cite{BraunMunzinger:1994xr,
BraunMunzinger:1995bp, BraunMunzinger:1999qy,
BraunMunzinger:2001ip}. At the same time among many measurements
of the hadron yields from heavy ion collision, those of the light
nuclei are remarkable. This is so because light nuclei are
composed of nucleons produced at chemical freeze-out and hence
expected to be produced some time after the chemical freeze-out.
On the other hand, the yields of light nuclei are explainable in
the statistical hadronization model with the same conditions at
which normal hadrons are produced. Another puzzle is the
production yield ratio of $K^*/K$, which is overestimated by the
statistical model \cite{BraunMunzinger:2001ip}. Moreover, the
ratio is smaller at Large Hadron Collider (LHC) than at
Relativistic Heavy Ion Collidier (RHIC) \cite{Adams:2004ep,
Abelev:2014uua}.

In general, production yields of hadronic molecules, light nuclei
or resonances not only depend on the conditions at the chemical
freeze-out but also on their interactions with other hadrons
during the hadronic stage. Specifically, for the light nuclei or
hadronic molecules that are bound, the yields are affected by the
freeze-out condition of the respective particle in the hadronic
matter. For resonances that decay into daughter particles, the
freeze-out point of the daughter particles will be important as
the resonances are reconstructed from the observed daughter
particles. In other words, investigation on the yields of hadronic
molecules or resonances may result in understanding of how long
the hadronic stage lasts in heavy ion collisions, or of when the
kinetic freeze-out of various particles takes place. In this
letter, we focus on the freeze-out conditions and discuss natural
consequences that follow from applying these conditions to
understand the production yields of resonances and hadronic
molecules.

Kinetic freeze-out of a particle species $i$ from the matter
occurs when its scattering rate $\tau^i_{scatt}$ becomes larger
than the expansion rate of the system $\tau_{exp}$
\cite{Bondorf:1978kz}. Hence, in an expanding system of
interacting particles, freeze-out takes place when
\begin{equation}
\tau_{exp}=\frac{1}{\partial\cdot u}=\tau_{scatt}^i=\frac{1}
{\sum_j\langle\sigma_{ij} v_{ij}\rangle n_j}, \label{criterion}
\end{equation}
with $\langle\sigma_{ij} v_{ij}\rangle$ being the thermally
averaged cross section times the relative velocity between
particle species $i$ and $j$, $n_j$ the density of particle $j$,
and $u$ the expansion velocity of the system.

The expansion time $\tau_{exp}$ can be approximated to the ratio
of the fireball volume $V$ to the change of that in time,
$V/(dV/dt)$, which can be further reduced to $R/(3dR/dt)$ for the
spherically expanding fireball with its radius $R$. Let us for
simplicity further assume that the system is composed of one
species only and that the cross section is velocity independent.
The freeze-out condition in Eq. (\ref{criterion}) then becomes
\begin{equation}
\frac{R}{3dR/dt}=\frac{1}{n\sigma\langle v\rangle}. \label{criteria1}
\end{equation}
In general the relation between $dR/dt$ and $\langle v \rangle$ is
not universal; particularly so when there is a flow. However,
assuming that the rate of change in the radius is close to the
average velocity of the particles in the system, the condition for
the kinetic freeze-out becomes
\begin{equation}
\frac{N}{R_{fo}^2}=\frac{4\pi}{\sigma_{fo}}, \label{sigmatoR2}
\end{equation}
where a subscript $_{fo}$ stands for physical quantities at
kinetic freeze-out and $N$ is the total number of particles. We
see that the two dimensional density determines the condition for
freeze-out, because the transverse total cross section determines
whether the particle interacts with the medium when it escapes
from the medium.

On the other hand, the three dimensional density at freeze-out
then goes as
\begin{equation}
\frac{N}{R_{fo}^3}= \bigg(\frac{4 \pi}{\sigma_{fo}}
\bigg)^{3/2}\frac{1}{N^{1/2}}. \label{three-density}
\end{equation}
This suggests that for higher collision energies and/or when the
initial temperature and/or the number of particles increases, the
three dimensional density at which freeze-out takes places becomes
smaller. Let us discuss the relevance of this result in
understanding the $K^*$ and/or resonance production in heavy ion
collisions.

It has been known that the experimentally measured yield of $K^*$
mesons does not agree with the statistical hadronization model
prediction \cite{BraunMunzinger:2001ip, Adams:2004ep,
Abelev:2014uua}. Due to the short lifetime of the $K^*$ meson
compared to the lifespan of the hadronic stage in heavy ion
collisions, $K^*$ mesons not only participate in the hadronic
interactions with light mesons but also decay to and reform from
kaons and pions during the hadronic stage. As the daughter
particles of $K^*$ mesons are subject to re-scatter as well in the
hadronic medium, the measured $K^*$ meson reconstructed from an
invariant mass analysis of the daughter particles will reflect its
number when the daughter particles freeze out from the medium.

In order to investigate the hadronic effects on the $K^*$ meson
abundance in the hadronic medium in general, we consider
simplified rate equations for the abundances of both $K^*$ and $K$
mesons during the hadronic stage,
\begin{eqnarray}
&&\frac{dN_{K^*}(\tau)}{d\tau}=\frac{1}{\tau^K_{scatt}}
N_{K}(\tau)-\frac{1}{\tau^{K^*}_{scatt}}N_{K^*}(\tau), \nonumber \\
&&\frac{dN_K(\tau)}{d\tau}=\frac{1}{\tau^{K^*}_{scatt}}N_{K^*}(\tau)-
\frac{1}{\tau^K_{scatt}}N_{K}(\tau), \label{NKvKsrateSimple}
\end{eqnarray}
with $1/\tau^{K^*}_{scatt}=\sum_{i}\langle\sigma_{K^* i}v_{K^*
i}\rangle n_i$, and $1/\tau^K_{scatt}=\sum_j\langle\sigma_{K
j}v_{K j}\rangle n_j$, Eq. (\ref{criterion}). Here $i$, $j$ stands
for mostly the light mesons such as pions and $\rho$ mesons, i.e.,
$1/\tau^{K^*}_{scatt}=\langle\sigma_{K^*\rho\to K\pi}v_{K^*\rho}
\rangle n_{\rho}+\langle\sigma_{K^*\pi\to K\rho}v_{K^*\pi} \rangle
n_{\pi}+\langle\Gamma_{K^*}\rangle$,
$1/\tau^K_{scatt}=\langle\sigma_{K\pi\to K^*\rho}v_{K\pi} \rangle
n_{\pi}+\langle\sigma_{K\rho\to K\rho}v_{K\rho} \rangle
n_{\rho}+\langle\sigma_{K\pi\to K^*}v_{K\pi}\rangle n_{\pi}$ with
$\langle\Gamma_{K^*}\rangle$ being the thermally averaged decay
width of the $K^*$ meson \cite{Cho:2015qca}. Here we do not
consider non-linear terms originated from the interaction between
$K^*$ mesons or kaons, like $K\bar{K}\to\rho\pi$.

While the exact solution of Eq. (\ref{NKvKsrateSimple}) including
the non-linear terms was obtained in Ref. \cite{Cho:2015qca}, the
essential physics can be understood in the following simple limit.
When the thermal cross sections and densities of light mesons are
independent of time, the following analytic solution for the yield
ratio between $K^*$ mesons and kaons, $R=N_{K^*}/(N_K+N_{K^*})$
can be obtained from the above coupled differential equations,
\begin{eqnarray}
R(\tau)=R_0+\Big(\frac{N_{K^*}^0}
{N^0}-\frac{\tau_{scatt}}{\tau_{scatt}^K}\Big)e^{-\frac{\tau-\tau_h}{
\tau_{scatt}}}. \label{ratio_tau}
\end{eqnarray}
with $R_0=\tau_{scatt}/\tau_{scatt}^K$ and
$\tau_{scatt}=\tau_{scatt}^{K^*}\tau_{scatt}^K/(\tau_{scatt}^
{K^*}+\tau_{scatt}^K)$. The initial yields for both hadrons are
assumed to be $N_K^0$ and $N_{K^*}^0$ and $N^0=N_K^0+N_{K^*}^0$.
In Eq. (\ref{ratio_tau}) the first term $R_0$ represents the
abundance ratio of $K^*$ mesons to kaon at the equilibrium
temperature, while the second term through $\tau_{scatt}$
determines the rate at which the new equilibrium number is
reached. If the $\tau_{scatt}$ is small, the abundance reaches the
equilibrium ratio instantly.

It has been found that the actual variation of the abundance ratio
of $K^*$ mesons to kaons during the hadronic stage shows similar
features as the solution, Eq. (\ref{ratio_tau}) but with a
decreasing equilibrium ratio $R_0$ due to the decreasing
background temperature \cite{Cho:2015qca}. Since the transient
term in the yield ratio $R(\tau)$ plays a negligible role at a
later time during the hadronic interaction stage, the equilibrium
ratio, $R_0=\tau_{scatt}/\tau_{scatt}^K$ that reflects the thermal
$K^*/K$ ratio of the hadron gas at the background temperature,
mainly determines the  ratio of $K^*$ mesons to kaons towards the
end of the hadronic stage.

The $R_0=\tau_{scatt}/\tau_{scatt}^K$ decreases as the system
expands and the temperature of the hadron gas decreases.
Therefore, the final ratio between $K^*$ and $K$ mesons may
reflect the condition at the last stage of the hadronic effects on
$K^*$ and $K$ mesons, the kinetic freeze-out temperature. The
$K^*$ meson is measured in experiment by reconstructing the
invariant mass of $K$ and $\pi$ mesons. If the freeze-out times
for $K^*$, $K$ and $\pi$ mesons are same, then the measured
$K^*/K$ yield will reflect the temperature at the single
freeze-out temperature. However, if the freeze-out time of $\pi$
mesons is later than those of $K^*$ and $K$ mesons, due to the
larger cross section with the medium, the part of $K^*$ mesons
will be lost in the reconstruction, because the additional
scattering of pions (after the freeze-out of $K^*$ mesons) will
wash out the origin of pions. Therefore, the experimentally
measured $K^*/K$ ratio will reflect the freeze-out temperature of
$\pi$ mesons.

The reduction of the $K^*$ meson yield from the statistical
hadronization model expectation is found to be more significant at
LHC than at RHIC. As discussed before, the ratio of $K^*$ to $K$
mesons reflects the condition at the last stage of the $K^*$ meson
interaction in the hadronic medium so that the reduction of the
ratio between $K^*$ mesons and kaons at LHC originates from the
lower kinetic freeze-out temperature of the system at LHC compared
to that at RHIC.

The idea of the lower kinetic freeze-out temperature at LHC
compared to that at RHIC becomes evident if we recall the density
of particles at the kinetic freeze-out in heavy ion collisions,
Eq. (\ref{three-density}). That is, the density at freeze-out
becomes smaller when the number of initially produced particles
increases. In the following, we will demonstrate that the
freeze-out temperature for pions are lower at LHC than at RHIC by
explicitly applying the condition in Eq. (\ref{criterion}) to both
cases.

\begin{figure}[!h]
\begin{center}
\includegraphics[width=0.50\textwidth]{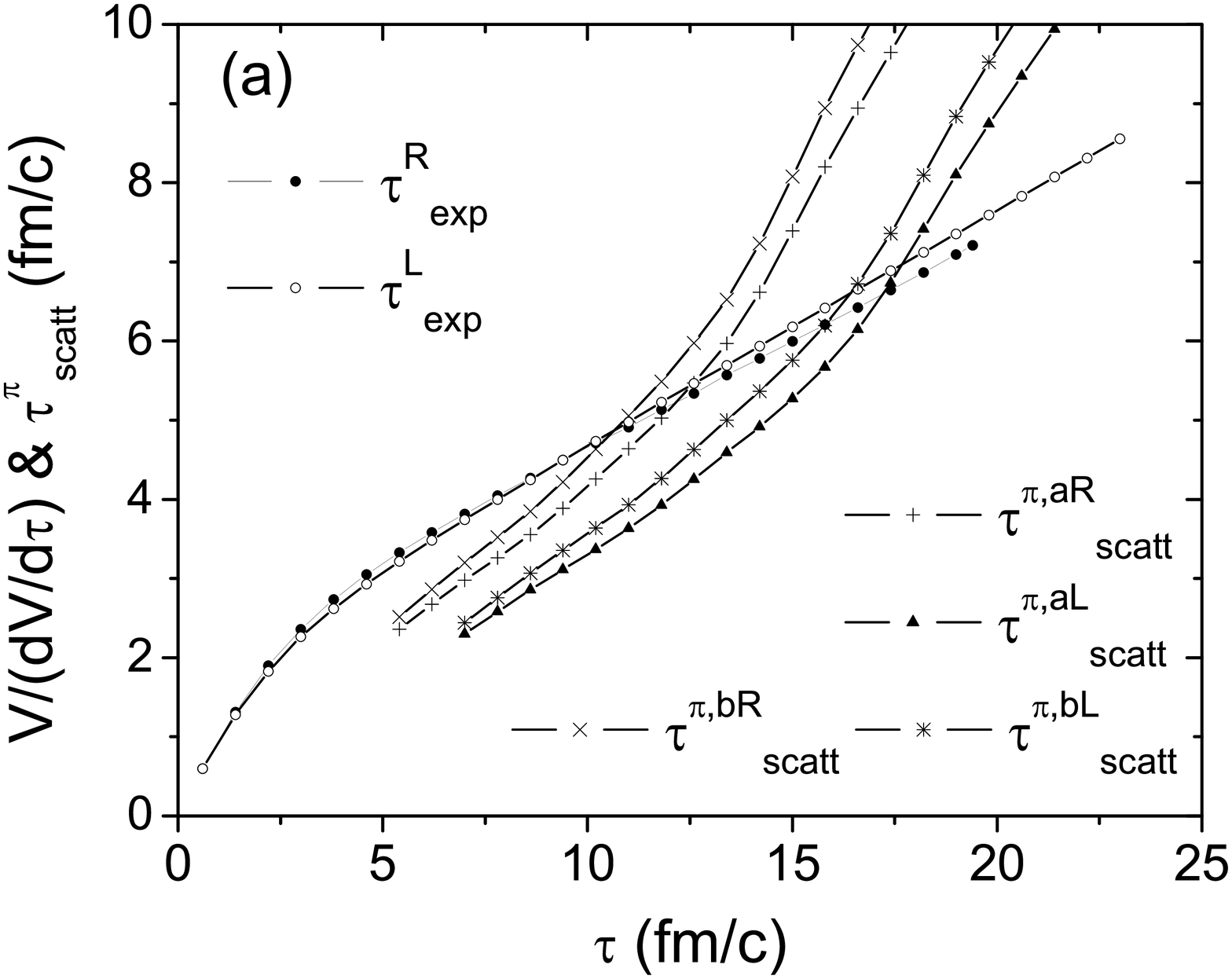}
\includegraphics[width=0.50\textwidth]{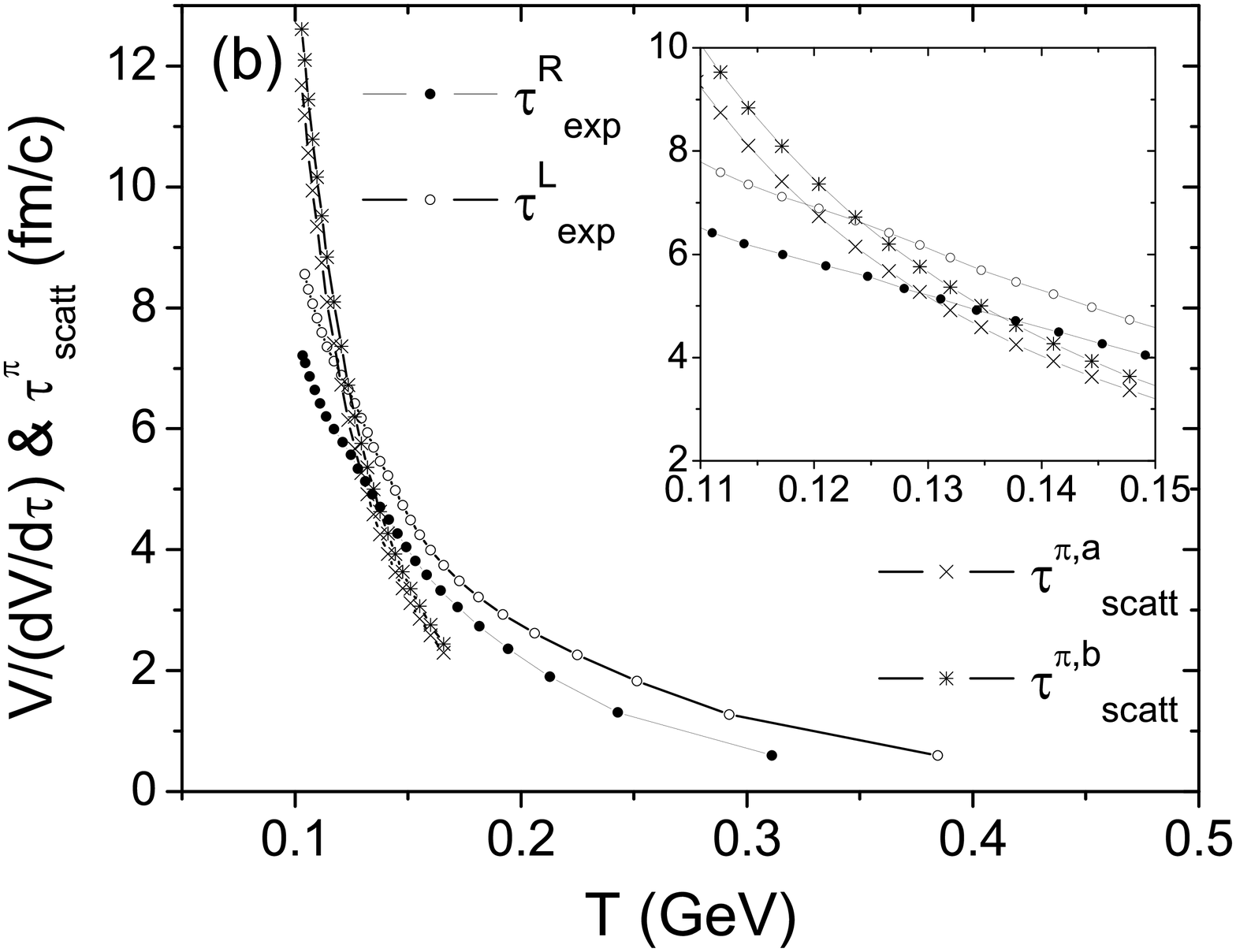}
\end{center}
\caption{Variations of the $\tau_{scatt}^\pi$ and the expansion
time of the system, $\tau_{exp}=V/(dV/dt)$ at RHIC and LHC both in
time (a) and temperature (b) during the hadronic stage.}
\label{tau_scat to tau_exp}
\end{figure}

First, we evaluate the expansion time of the system
$\tau_{exp}=V/(dV/dt)$ for both at RHIC and at LHC using
hydrodynamics simulations. Hydrodynamic equations are expressed as
$\partial_\mu T^{\mu\nu}=0$, where the energy-momentum tensor
$T^{\mu\nu}=(e+p)u^\mu u^\nu-pg^{\mu\nu}$ with $e$, $p$ and
$u^\mu$ being, respectively, energy density, pressure, and
4-velocity of flow. Using $u_\nu\partial_\mu T^{\mu\nu}=0$ and the
first law of thermodynamics for infinitely small volume of
cell~\cite{Biro:1981es},
\begin{eqnarray}
\partial_\mu(e u^\mu)=T\partial_\mu(su^\mu)-p(\partial_\mu u^\mu),
\end{eqnarray}
with $T$ and $s$ being temperature and entropy density, we can
derive the entropy conservation, $\partial_\mu (su^\mu)=0$. For
simplicity, we assume boost-invariance and consider central
collisions, that is, symmetric expansion in transverse plane. Then
there are only two independent hydrodynamic
equations~\cite{Heinz:2005bw}:
\begin{eqnarray}
\frac{1}{\tau}\partial_\tau (\tau T^{\tau\tau})+\frac{1}{r}\partial_r (r T^{\tau r})=\frac{p}{\tau},\label{energy}\\
\frac{1}{\tau}\partial_\tau (\tau su^\tau)+\frac{1}{r}\partial_r (r su^r)=0,\label{entropy}
\end{eqnarray}
where $\tau=\sqrt{t^2-z^2}$ and $r=\sqrt{x^2+y^2}$. Integrating
the above equations over transverse plane, we have
\begin{eqnarray}
\frac{1}{\tau}\partial_\tau (\tau \int dA T^{\tau\tau})=\frac{1}{\tau}\int dA p,\\
\frac{1}{\tau}\partial_\tau (\tau \int dA su^\tau)=0,
\end{eqnarray}
where $dA=2\pi rdr$. We note that the second terms in the
left-hand side of Eq.~(\ref{energy}) and (\ref{entropy}) disappear
due to boundary condition. Here we make the assumption that
nuclear matter has definite boundary and $e$, $s$, and $p$ are
uniform inside. Then we have the following simple equations:
\begin{equation}
\frac{1}{\tau}\partial_\tau (\tau A \langle T^{\tau\tau}\rangle)=
\frac{1}{\tau} pA, \quad\frac{1}{\tau}\partial_\tau (\tau A s
\langle u^\tau \rangle)=0\label{entropy2},
\end{equation}
where $A=\pi R^2$ with $R$ being the radius of nuclear matter and
$\langle T^{\tau\tau}\rangle=\int dA T^{\tau\tau}/A=(e+p)\langle
\gamma_r^2\rangle-p, \langle u^\tau \rangle=\langle \gamma_r
\rangle,$ with $\gamma_r=1/\sqrt{1-v_r^2}$ and $v_r$ being radial
velocity. Assuming that the radial flow velocity is a linear
function of the radial distance from the center, i.e., $\gamma_r
v_r=\gamma_R \dot{R}(r/R)$, where $\dot{R}=\partial R/\partial
\tau$ and $\gamma_R=1/\sqrt{1-\dot{R}^2}$,
\begin{equation}
\langle\gamma_r^2\rangle=1+\frac{\gamma_R^2 \dot{R}^2}{2},
\quad\langle\gamma_r\rangle=\frac{2}{3\gamma_R^2
\dot{R}^2}\left(\gamma_R^3-1\right). \label{gamma}
\end{equation}
We then numerically solve Eq.~(\ref{entropy2}) by using the
lattice equation of state~\cite{Song:2010fk, Borsanyi:2010cj}. The
initial thermalization time for hydrodynamic simulations is
assumed 0.5 fm/c, and the initial radius is given by the
transverse area where local temperature is above 150 MeV. Though
the hydrodynamic approach is marginal in hadron gas phase, it has
successfully reproduced abundant experimental data from
relativistic heavy-ion
collisions~\cite{Kolb:2000sd,Schenke:2010nt}. We show the results
for LHC and RHIC as a function of both the proper time and the
temperature in Fig. \ref{tau_scat to tau_exp}.

Second, we consider $1/\tau_{scatt}\approx\sigma_{\pi}\langle
v\rangle n_\pi$. Taking the total cross section of the pion in the
hadronic medium to be 40 mb, $\langle v\rangle$=0.7$c$, and
assuming that pions are in thermal equilibrium during the hadronic
stage, $n_{\pi}(T)\approx \frac{g_{\pi}}{2\pi^2}m_{\pi}^2 T
K_2(m_{\pi}/T)$, with $g_{\pi}$ and $m_\pi$ being the degeneracy
and mass of pions, respectively, and $K_2$ the modified Bessel
function of the second kind, we obtain $\tau_{scatt}^{\pi,a}$
shown in Fig. \ref{tau_scat to tau_exp}. In addition, we consider
a numerically evaluated pion scattering time \cite{Hung:1997du}
parameterized as $1/\tau_{scatt}^{\pi}$=(59.5
fm$^{-1})(T/1GeV)^{3.45}$ \cite{Heinz:2007in}. In Fig.
\ref{tau_scat to tau_exp} we show the above parameterized
scattering time multiplied by the phenomenological parameter
$\xi=0.295$ \cite{Heinz:2007in}, denoted by
$\tau_{scatt}^{\pi,b}$. When evaluating pion scattering times, we
have used the temperature as a function of proper time, $T(\tau)$
obtained from hydrodynamics simulations.

As shown in Fig. \ref{tau_scat to tau_exp} (a), both pion
scattering times for RHIC, $\tau_{scatt}^{\pi,aR}$ and
$\tau_{scatt}^{\pi,bR}$ cross the expansion time of the system
earlier than that for LHC, $\tau_{scatt}^{\pi,aL}$ and
$\tau_{scatt}^{\pi,bL}$, implying that the kinetic freeze-out
temperature at RHIC is higher than that at LHC; Fig. \ref{tau_scat
to tau_exp} (b) clearly demonstrates this. The line of the pion
scattering time crosses the expansion velocity of the system at
RHIC before that at LHC. We read in the inset of Fig.
\ref{tau_scat to tau_exp} (b) that the kinetic freeze-out occurs
at about 130 MeV at RHIC, and at about 120 MeV at LHC from
$\tau_{scatt}^{\pi,a}$, or at about 135 MeV at RHIC, and at about
125 MeV at LHC from $\tau_{scatt}^{\pi,b}$. These values are
consistent with the temperatures needed to explain the
experimentally measured $K^*/K$ ratio at RHIC and LHC with the
solutions of Eq. (\ref{NKvKsrateSimple}) \cite{Cho:2015qca}.

The above argument applies to any resonances or hadronic molecules
that decay to certain daughter particles. It should be noted also
that the yields of the resonances or hadronic molecules will
decrease when the cross section of the daughter particles with the
medium increases. At the same time, for strongly bound states such
as the light nuclei, the freeze-out temperature will be determined
by the cross section of the bound state itself with the medium.

Let us consider the coalescence production of the deuteron during
the hadronic stage in heavy ion collisions. In the same model used
in the evaluation of hadronic molecule states \cite{Lee:2015fsa},
the yield of deuteron is given by,
\begin{equation}
N_D=\frac{g_D}{g_N^2}\frac{N_N^2}{V_{fo}}\frac{(4\pi
a_D^2)^{3/2}}{1+2m_N T_{fo}a_D^2}, \label{CoalD}
\end{equation}
where $g_D$ and $g_N$ are the degeneracy of the deuteron and
nucleon, respectively, $N_N$ the number of nucleons, and $m_N$ the
mass of the nucleon. We represent the freeze-out volume at
hadronization as $V_{fo}$ and the hadronization temperature
$T_{fo}$. $a_D$ is the radius of the deuteron. Assuming that the
size of the final bound state is related to the cross section of
the state with the medium,  the above can be written again as,
\begin{eqnarray}
&& N_D=\frac{g_D}{g_N^2}N_N^2\frac{3}{4\pi}\Big(4\frac{\sigma_D}
{R_{fo}^2}\Big)^{3/2}\frac{1}{1+2m_N T_{fo} a_D^2} \nonumber \\
&& \quad~=\frac{g_D}{g_N^2}N_N^2\frac{48\pi^{1/2}}{
N_\pi^{3/2}}\frac{1}{1+2m_N T_{fo} a_D^2},
\end{eqnarray}
where we have assumed that $V_{fo}=4\pi/3 R_{fo}^3$ and
$\sigma_{fo}=\pi a_D^2=\sigma_D$, the cross section of the
deuteron, and have used the condition at the freeze-out, the
relation between the cross section of the deuteron $\sigma_{fo}$
and the freeze-out radius $R_{fo}$, Eq. (\ref{sigmatoR2}) with
pions as the dominant constituents of the medium. As we see, the
coalescence production of the deuteron is almost independent of
the size of the deuteron and only depends on the number of
constituents of the medium. When the cross section is large, it
freezes out later. Assuming that the number of ground state
constituents do not change much during the hadronic phase, one
could explain why the observed production yield of the deuteron is
well understood in the statistical hadronization model .

We have shown that the yield ratio of $K^*/K$ from heavy ion
collisions at RHIC and at LHC can be understood by applying the
kinetic freeze-out condition to the daughter particles. Since the
ratio is manly determined from the relative interaction ratio,
$\tau_{scatt}/\tau_{scatt}^K$, which reflects temperature of the
system, it is possible to infer the condition at the kinetic
freeze-out. The larger suppression of the ratio at LHC than at
RHIC compared to the expectations from statistical model based on
chemical freeze-out parameters reflects the lower temperature at
LHC as demonstrated explicitly by applying the freeze-out
condition to hydrodynamic calculations.

Furthermore, we have pointed out that for the light nuclei or
hadronic molecules that are bound, the yields are affected by the
freeze-out condition of the respective particle with the hadronic
matter. This led to the observation that the deuteron production
yields are independent of the size of deuteron, and depend only on
the number of ground state constituents. We therefore suggest that
studying production of various hadrons such as hadronic molecules,
resonances and light hadrons not only provides us chances to
understand hadron formation in heavy ion collisions but also gives
us valuable information on the evolution of the entire system
itself.

\textit{Acknowledgements} We thank Che Ming Ko for useful
discussions. S. H. Lee was supported by the Korea National
Research Foundation under the grant number KRF-2011-0020333 and
KRF-2011-0030621. S. Cho was supported by 2015 Research Grant from
Kangwon National University.

\end{document}